
%
%
%
%
%
%
%
%
%
%
\def\unlockat{\catcode`\@=11}
\def\lockat{\catcode`\@=12}
\unlockat
\def\d@f@ult{} \newif\ifamsfonts \newif\ifafour
%
%

\font\twelverm=cmr12
\font\ninerm=cmr9
\font\sixrm=cmr6
\font\fourteenbf=cmbx12 scaled\magstep1
\font\twelvebf=cmbx12
\font\ninebf=cmbx9
\font\sixbf=cmbx6
\font\fourteeni=cmmi12 scaled\magstep1      \skewchar\fourteeni='177
\font\twelvei=cmmi12                        \skewchar\twelvei='177
\font\ninei=cmmi9                           \skewchar\ninei='177
\font\sixi=cmmi6                            \skewchar\sixi='177
\font\fourteensy=cmsy10 scaled\magstep2     \skewchar\fourteensy='60
\font\twelvesy=cmsy10 scaled\magstep1       \skewchar\twelvesy='60
\font\ninesy=cmsy9                          \skewchar\ninesy='60
\font\sixsy=cmsy6                           \skewchar\sixsy='60
\font\fourteenex=cmex10 scaled\magstep2
\font\twelveex=cmex10 scaled\magstep1

\ifamsfonts
   \font\ninex=cmex9
   
   \font\sixex=cmex7 at 6pt
   
\else
   \font\ninex=cmex10 at 9pt
   
   \font\sixex=cmex10 at 6pt
   
\fi
\font\fourteensl=cmsl10 scaled\magstep2
\font\twelvesl=cmsl10 scaled\magstep1

\font\sevensl=cmsl10 at 7pt
\font\sixsl=cmsl10 at 6pt

\font\fourteenit=cmti12 scaled\magstep1
\font\twelveit=cmti12

\font\fourteentt=cmtt12 scaled\magstep1
\font\twelvett=cmtt12
\font\fourteencp=cmcsc10 scaled\magstep2
\font\twelvecp=cmcsc10 scaled\magstep1

\ifamsfonts
   
\else
   
\fi
\newfam\cpfam
\font\fourteenss=cmss12 scaled\magstep1
\font\twelvess=cmss12
\font\tenss=cmss10
\font\niness=cmss9

\font\sevenss=cmss8 at 7pt
\font\sixss=cmss8 at 6pt
\newfam\ssfam
\newfam\msafam \newfam\msbfam \newfam\eufam
\ifamsfonts
 \font\fourteenmsa=msam10 scaled\magstep2
 \font\twelvemsa=msam10 scaled\magstep1
 \font\tenmsa=msam10
 \font\ninemsa=msam9
 \font\sevenmsa=msam7
 \font\sixmsa=msam6
 \font\fourteenmsb=msbm10 scaled\magstep2
 \font\twelvemsb=msbm10 scaled\magstep1
 \font\tenmsb=msbm10
 \font\ninemsb=msbm9
 \font\sevenmsb=msbm7
 \font\sixmsb=msbm6
 \font\fourteeneu=eufm10 scaled\magstep2
 \font\twelveeu=eufm10 scaled\magstep1
 \font\teneu=eufm10
 \font\nineeu=eufm9
 
 \font\seveneu=eufm7
 \font\sixeu=eufm6
 \def\hexnumber@#1{\ifnum#1<10 \number#1\else
  \ifnum#1=10 A\else\ifnum#1=11 B\else\ifnum#1=12 C\else
  \ifnum#1=13 D\else\ifnum#1=14 E\else\ifnum#1=15 F\fi\fi\fi\fi\fi\fi\fi}
 \def\hexmsa{\hexnumber@\msafam}
 \def\hexmsb{\hexnumber@\msbfam} 
\fi
\newdimen\b@gheight             \b@gheight=12pt
\newcount\f@ntkey               \f@ntkey=0
\def\f@m{\afterassignment\samef@nt\f@ntkey=}
\def\samef@nt{\fam=\f@ntkey \the\textfont\f@ntkey\relax}
\def\rm{\f@m0 }
\def\mit{\f@m1 }
\def\cal{\f@m2 }
\def\it{\f@m\itfam}
\def\sl{\f@m\slfam}
\def\bf{\f@m\bffam}
\def\tt{\f@m\ttfam}
\def\caps{\f@m\cpfam}
\def\ssf{\f@m\ssfam}
\ifamsfonts
 \def\msa{\f@m\msafam}
 \def\msb{\f@m\msbfam} 
 \def\eu{\f@m\eufam}
\else
  \let\eu=\bf
\fi
\def\fourteenpoint{\relax
    \textfont0=\fourteencp          \scriptfont0=\tenrm
      \scriptscriptfont0=\sevenrm
    \textfont1=\fourteeni           \scriptfont1=\teni
      \scriptscriptfont1=\seveni
    \textfont2=\fourteensy          \scriptfont2=\tensy
      \scriptscriptfont2=\sevensy
    \textfont3=\fourteenex          \scriptfont3=\twelveex
      \scriptscriptfont3=\tenex
    \textfont\itfam=\fourteenit     \scriptfont\itfam=\tenit
    \textfont\slfam=\fourteensl     \scriptfont\slfam=\tensl
      \scriptscriptfont\slfam=\sevensl
    \textfont\bffam=\fourteenbf     \scriptfont\bffam=\tenbf
      \scriptscriptfont\bffam=\sevenbf
    \textfont\ttfam=\fourteentt
    \textfont\cpfam=\fourteencp
    \textfont\ssfam=\fourteenss     \scriptfont\ssfam=\tenss
      \scriptscriptfont\ssfam=\sevenss
    \ifamsfonts
       \textfont\msafam=\fourteenmsa     \scriptfont\msafam=\tenmsa
         \scriptscriptfont\msafam=\sevenmsa
       \textfont\msbfam=\fourteenmsb     \scriptfont\msbfam=\tenmsb
         \scriptscriptfont\msbfam=\sevenmsb
       \textfont\eufam=\fourteeneu     \scriptfont\eufam=\teneu
         \scriptscriptfont\eufam=\seveneu \fi
    \samef@nt
    \b@gheight=14pt
    \setbox\strutbox=\hbox{\vrule height 0.85\b@gheight
                                depth 0.35\b@gheight width\z@ }}
\def\twelvepoint{\relax
    \textfont0=\twelverm          \scriptfont0=\ninerm
      \scriptscriptfont0=\sixrm
    \textfont1=\twelvei           \scriptfont1=\ninei
      \scriptscriptfont1=\sixi
    \textfont2=\twelvesy           \scriptfont2=\ninesy
      \scriptscriptfont2=\sixsy
    \textfont3=\twelveex          \scriptfont3=\ninex
      \scriptscriptfont3=\sixex
    \textfont\itfam=\twelveit    
    \textfont\slfam=\twelvesl    
      \scriptscriptfont\slfam=\sixsl
    \textfont\bffam=\twelvebf     \scriptfont\bffam=\ninebf
      \scriptscriptfont\bffam=\sixbf
    \textfont\ttfam=\twelvett
    \textfont\cpfam=\twelvecp
    \textfont\ssfam=\twelvess     \scriptfont\ssfam=\niness
      \scriptscriptfont\ssfam=\sixss
    \ifamsfonts
       \textfont\msafam=\twelvemsa     \scriptfont\msafam=\ninemsa
         \scriptscriptfont\msafam=\sixmsa
       \textfont\msbfam=\twelvemsb     \scriptfont\msbfam=\ninemsb
         \scriptscriptfont\msbfam=\sixmsb
       \textfont\eufam=\twelveeu     \scriptfont\eufam=\nineeu
         \scriptscriptfont\eufam=\sixeu \fi
    \samef@nt
    \b@gheight=12pt
    \setbox\strutbox=\hbox{\vrule height 0.85\b@gheight
                                depth 0.35\b@gheight width\z@ }}
\twelvepoint
%
%
\baselineskip = 15pt plus 0.2pt minus 0.1pt 
\lineskip = 1.5pt plus 0.1pt minus 0.1pt
\lineskiplimit = 1.5pt
\parskip = 6pt plus 2pt minus 1pt
\interlinepenalty=50
\interfootnotelinepenalty=5000
\predisplaypenalty=9000
\postdisplaypenalty=500
\hfuzz=1pt
\vfuzz=0.2pt
\dimen\footins=24 truecm 
\ifafour
 \hsize=16cm \vsize=22cm
\else
 \hsize=6.5in \vsize=9in
\fi
%
%
\skip\footins=\medskipamount
\newcount\fnotenumber
\def\clearfnotenumber{\fnotenumber=0} \clearfnotenumber
\def\fnote{\global\advance\fnotenumber by1 \generatefootsymbol
 \footnote{$^{\footsymbol}$}}
\def\fd@f#1 {\xdef\footsymbol{\mathchar"#1 }}
\def\generatefootsymbol{\iffrontpage\ifcase\fnotenumber
\or \fd@f 279 \or \fd@f 27A \or \fd@f 278 \or \fd@f 27B
\else  \fd@f 13F \fi
\else\xdef\footsymbol{\the\fnotenumber}\fi}
%
%
\newcount\secnumber \newcount\appnumber
\def\clearappnumber{\appnumber=64} \def\clearsecnumber{\secnumber=0}
\clearsecnumber \clearappnumber
\newif\ifs@c 
\newif\ifs@cd 
\s@cdtrue 
\def\unsectioned{\s@cdfalse\let\section=\subsection}
\newskip\sectionskip         \sectionskip=\medskipamount
\newskip\headskip            \headskip=8pt plus 3pt minus 3pt
\newdimen\sectionminspace    \sectionminspace=10pc
\def\Titlestyle#1{\par\begingroup \interlinepenalty=9999
     \leftskip=0.02\hsize plus 0.23\hsize minus 0.02\hsize
     \rightskip=\leftskip \parfillskip=0pt
     \advance\baselineskip by 0.5\baselineskip
     \hyphenpenalty=9000 \exhyphenpenalty=9000
     \tolerance=9999 \pretolerance=9000
     \spaceskip=0.333em \xspaceskip=0.5em
     \fourteenpoint
  \noindent #1\par\endgroup }
\def\titlestyle#1{\par\begingroup \interlinepenalty=9999
     \leftskip=0.02\hsize plus 0.23\hsize minus 0.02\hsize
     \rightskip=\leftskip \parfillskip=0pt
     \hyphenpenalty=9000 \exhyphenpenalty=9000
     \tolerance=9999 \pretolerance=9000
     \spaceskip=0.333em \xspaceskip=0.5em
     \fourteenpoint
   \noindent #1\par\endgroup }
\def\spacecheck#1{\dimen@=\pagegoal\advance\dimen@ by -\pagetotal
   \ifdim\dimen@<#1 \ifdim\dimen@>0pt \vfil\break \fi\fi}
\def\section#1{\cleareqnumber \s@ctrue \global\advance\secnumber by1
   \par \ifnum\the\lastpenalty=30000\else
   \penalty-200\vskip\sectionskip \spacecheck\sectionminspace\fi
   \noindent {\caps\enspace\S\the\secnumber\quad #1}\par
   \nobreak\vskip\headskip \penalty 30000 }
\def\undertext#1{\vtop{\hbox{#1}\kern 1pt \hrule}}
\def\subsection#1{\par
   \ifnum\the\lastpenalty=30000\else \penalty-100\smallskip
   \spacecheck\sectionminspace\fi
   \noindent\undertext{#1}\enspace \vadjust{\penalty5000}}

\def\appendix#1{\cleareqnumber \s@cfalse \global\advance\appnumber by1
   \par \ifnum\the\lastpenalty=30000\else
   \penalty-200\vskip\sectionskip \spacecheck\sectionminspace\fi
   \noindent {\caps\enspace Appendix \char\the\appnumber\quad #1}\par
   \nobreak\vskip\headskip \penalty 30000 }
\def\ack{\par\penalty-100\medskip \spacecheck\sectionminspace
   \line{\fourteencp\hfil ACKNOWLEDGEMENTS\hfil}%
\nobreak\vskip\headskip }
\def\refs{\begingroup \par\penalty-100\medskip \spacecheck\sectionminspace
   \line{\fourteencp\hfil REFERENCES\hfil}%
\nobreak\vskip\headskip \frenchspacing }
\def\endrefs{\par\endgroup}
%
%
\newif\iffrontpage \frontpagefalse
\headline={\hfil}
\footline={\iffrontpage\hfil\else \hss\twelverm
-- \folio\ --\hss \fi }
%
%
\newskip\frontpageskip \frontpageskip=12pt plus .5fil minus 2pt
\def\titlepage{\global\frontpagetrue\hrule height\z@ \relax
               \pubblock\relax }
\def\endtitlepage{\vfil\break\clearfnotenumber\frontpagefalse}
\def\title#1{\vskip\frontpageskip\Titlestyle{\caps #1}\vskip3\headskip}
\def\author#1{\vskip.5\frontpageskip\titlestyle{\caps #1}\nobreak}
\def\and{\par\kern 5pt \centerline{\sl and}}

\def\address#1{\par\kern 5pt\titlestyle{\it #1}}
\def\andaddress{\par\kern 5pt \centerline{\sl and} \address}

\def\abstract#1{\par\dimen@=\prevdepth \hrule height\z@ \prevdepth=\dimen@
   \vskip\frontpageskip\spacecheck\sectionminspace
   \centerline{\fourteencp ABSTRACT}\vskip\headskip
   {\noindent #1}}

\def\email#1{\fnote{\tentt e-mail: #1\hfill}}

%
%

%

%
\def\QMW{\address{%
   Department of Physics, Queen Mary and Westfield College\break
   Mile End Road, London E1 4NS, UK}}
%

%
%
\newcount\refnumber \def\clearrefnumber{\refnumber=0}  \clearrefnumber
\newwrite\R@fs                              
\immediate\openout\R@fs=\jobname.refs 
\def\closerefs{\immediate\closeout\R@fs} 
\def\refsout{\closerefs\refs
\unlockat
\input\jobname.refs
\lockat
\endrefs}
\def\refitem#1{\item{{\bf #1}}}
\def\ifundefined#1{\expandafter\ifx\csname#1\endcsname\relax}
\def\[#1]{\ifundefined{#1R@FNO}%
\global\advance\refnumber by1%
\expandafter\xdef\csname#1R@FNO\endcsname{[\the\refnumber]}%
\immediate\write\R@fs{\noexpand\refitem{\csname#1R@FNO\endcsname}%
\noexpand\csname#1R@F\endcsname}\fi{\bf \csname#1R@FNO\endcsname}}
\def\refdef[#1]#2{\expandafter\gdef\csname#1R@F\endcsname{{#2}}}
%
%
\newcount\eqnumber \def\cleareqnumber{\eqnumber=0}
\newif\ifal@gn \al@gnfalse  
\def\veqnalign#1{\al@gntrue \vbox{\eqalignno{#1}} \al@gnfalse}
\def\eqnalign#1{\al@gntrue \eqalignno{#1} \al@gnfalse}
\def\(#1){\relax%
\ifundefined{#1@Q}
 \global\advance\eqnumber by1
 \ifs@cd
  \ifs@c
   \expandafter\xdef\csname#1@Q\endcsname{{%
\noexpand\rm(\the\secnumber .\the\eqnumber)}}
  \else
   \expandafter\xdef\csname#1@Q\endcsname{{%
\noexpand\rm(\char\the\appnumber .\the\eqnumber)}}
  \fi
 \else
  \expandafter\xdef\csname#1@Q\endcsname{{\noexpand\rm(\the\eqnumber)}}
 \fi
 \ifal@gn
    & \csname#1@Q\endcsname
 \else
    \eqno \csname#1@Q\endcsname
 \fi
\else%
\csname#1@Q\endcsname\fi\global\let\@Q=\relax}
%
%
\newif\ifm@thstyle \m@thstylefalse
\def\mathstyle{\m@thstyletrue}
\def\proclaim#1#2\par{\smallbreak\begingroup
\advance\baselineskip by -0.25\baselineskip%
\advance\belowdisplayskip by -0.35\belowdisplayskip%
\advance\abovedisplayskip by -0.35\abovedisplayskip%
    \noindent{\caps#1.\enspace}{#2}\par\endgroup%
\smallbreak}
\def\m@kem@th<#1>#2#3{%
\ifm@thstyle \global\advance\eqnumber by1
 \ifs@cd
  \ifs@c
   \expandafter\xdef\csname#1\endcsname{{%
\noexpand #2\ \the\secnumber .\the\eqnumber}}
  \else
   \expandafter\xdef\csname#1\endcsname{{%
\noexpand #2\ \char\the\appnumber .\the\eqnumber}}
  \fi
 \else
  \expandafter\xdef\csname#1\endcsname{{\noexpand #2\ \the\eqnumber}}
 \fi
 \proclaim{\csname#1\endcsname}{#3}
\else
 \proclaim{#2}{#3}
\fi}
\def\Thm<#1>#2{\m@kem@th<#1M@TH>{Theorem}{\sl#2}}
\def\Prop<#1>#2{\m@kem@th<#1M@TH>{Proposition}{\sl#2}}
\def\Def<#1>#2{\m@kem@th<#1M@TH>{Definition}{\rm#2}}
\def\Lem<#1>#2{\m@kem@th<#1M@TH>{Lemma}{\sl#2}}
\def\Cor<#1>#2{\m@kem@th<#1M@TH>{Corollary}{\sl#2}}
\def\Conj<#1>#2{\m@kem@th<#1M@TH>{Conjecture}{\sl#2}}
\def\Rmk<#1>#2{\m@kem@th<#1M@TH>{Remark}{\rm#2}}
\def\Exm<#1>#2{\m@kem@th<#1M@TH>{Example}{\rm#2}}
\def\Qry<#1>#2{\m@kem@th<#1M@TH>{Query}{\it#2}}
%
%

%
\def\<#1>{\csname#1M@TH\endcsname}
%
%
\def\ref#1{{\bf [#1]}}
%
%

\def\lapprox{\hbox{\lower3pt\hbox{$\buildrel<\over\sim$}}}
\def\gapprox{\hbox{\lower3pt\hbox{$\buildrel<\over\sim$}}}
\def\quotient#1#2{#1/\lower0pt\hbox{${#2}$}}
\def\fr#1/#2{\mathord{\hbox{${#1}\over{#2}$}}}
\ifamsfonts
 \mathchardef\empty="0\hexmsb3F 
 \mathchardef\lsemidir="2\hexmsb6E 
 \mathchardef\rsemidir="2\hexmsb6F 
\else
 \let\empty=\emptyset
 \def\lsemidir{\mathbin{\hbox{\hskip2pt\vrule height 5.7pt depth -.3pt
    width .25pt\hskip-2pt$\times$}}}
 \def\rsemidir{\mathbin{\hbox{$\times$\hskip-2pt\vrule height 5.7pt
    depth -.3pt width .25pt\hskip2pt}}}
\fi
%
%

%
%
%
%
\def\underrightarrow#1{\vtop{\ialign{##\crcr
      $\hfil\displaystyle{#1}\hfil$\crcr
      \noalign{\kern-\p@\nointerlineskip}
      \rightarrowfill\crcr}}} 
\def\underleftarrow#1{\vtop{\ialign{##\crcr
      $\hfil\displaystyle{#1}\hfil$\crcr
      \noalign{\kern-\p@\nointerlineskip}
      \leftarrowfill\crcr}}}  

%
%
%
%

\def\NPB#1#2#3{{\sl Nucl. Phys.} {\bf B#1} (#2) #3}

\def\JMP#1#2#3{{\sl J. Math. Phys.} {\bf #1} (#2) #3}

\lockat
%
%

\def\cH{\mathord{\cal H}}
\def\cF{\mathord{\cal F}}
\def\gg{{\mathord{\eu g}}}
\def\gh{{\mathord{\eu h}}}
\let\tensor=\otimes
\let\isom=\cong
\def\fr#1/#2{\mathord{\hbox{${#1}\over{#2}$}}}
\def\half{\fr1/2}
\def\contour#1#2{\mathop{\oint_{C_{#2}} {{d#1}\over{2\pi i}}}}
\def\cS{\mathord{\cal S}}
\def\cA{\mathord{\cal A}}
\let\d=\partial
\def\W{\mathord{\ssf W}}
\refdef[BerVa]{N. Berkovits and C. Vafa, {\it On the Uniqueness of
String Theory}, {\tt hep-th/9310170}.}
\refdef[Knapp]{A.W. Knapp, {\it Lie Groups, Lie Algebras, and
Cohomology}, Mathematical Notes, PUP (1988) and in particular section
VI.2}
\refdef[homol]{J.M. Figueroa-O'Farrill, \NPB{343}{1990}{450}.}
\refdef[uniw]{J.M. Figueroa-O'Farrill and E. Ramos,
\JMP{33}{1992}{833} (Erratum: {\it ibid.} {\bf 34} (1993) 887).}
\refdef[ss]{For an introduction to spectral sequences try R. Bott and
L. Tu, {\it Differential Forms in Algebraic Topology}, Springer
Verlag, 1982.}
\refdef[Sasha]{A. A. Voronov, {\it Semi-infinite Homological Algebra},
Preprint 1993.}
\overfullrule=0pt
\unsectioned
\def\pubblock{ \line{\hfil\rm QMW--PH--93--29}
               \line{\hfil\tt hep-th/9310200}
               \line{\hfil\rm October 1993}}
\titlepage
\title{On the Universal String Theory}
\author{Jos\'e~M.~Figueroa-O'Farrill
\email{jmf@strings1.ph.qmw.ac.uk}}
\QMW
\abstract{Very recently Berkovits and Vafa have argued that the
$N{=}0$ string is a particular choice of background of the $N{=}1$
string.  Under the assumption that the physical states of the $N{=}0$
string theory came essentially from the matter degrees of freedom,
they proved that the amplitudes of both string theories agree.  They
also conjectured that this should persist whatever the form of the
physical states.  The aim of this note is to prove that both theories
have the same spectrum of physical states without making any
assumption on the form of the physical states.  We also notice in
passing that this result is reminiscent of a well-known fact in the
theory of induced representations and we explore what repercussions
this may have in the search for the universal string theory.}
\endtitlepage

\section{Introduction}

In a very recent paper Berkovits and Vafa \[BerVa] have argued that
$N{=}0$ string theory can be thought of as a special background
configuration of $N{=}1$ string theory and that, in turn, any $N{=}1$
string theory can be seen as a special background configuration of
$N{=}2$ string theory.  More concretely, what was shown in \[BerVa] is
that the amplitudes of the $N{=}0$ string theory could be recovered
from amplitudes of the $N{=}1$ string theory, and similarly for the
other embedding.  This was proven under the assumption that the
physical states were given essentially by excitations of the matter
degrees of freedom.  This condition, while it is certainly true for
critical string theory (away from zero momentum) is not generally
true, as evinced for instance, by the rich spectrum of the noncritical
string theories with $c<1$.  In the first footnote of \[BerVa], it is
already pointed out that their results should still hold true
regardless of the form of the physical states.  It is the purpose of
this note to elaborate on this footnote and to prove that for the
first of these embeddings, the physical spectrum of both string
theories agree.  We find it convenient to paraphrase the results of
\[BerVa] in the language of BRST cohomology.  Let us consider the case
of $N{=}0$ strings.

Let $T_m(z)$ denote the energy-momentum tensor of any CFT with
$c{=}26$.  Let $\cH$ denote the Hilbert space of this CFT.  The
physical states of the string with background $\cH$ are given by the
BRST cohomology $H_{N{=}0}(\cH)$.  Consider now the CFT defined by a
fermionic BC system $(b_1,c_1)$ of weights $(\fr3/2,-\fr1/2)$ and let
$\cF$ denote its Hilbert space.  It carries a representation of the
Virasoro algebra with $c{=}{-}11$.  More is true however and, as shown
in \[BerVa], $\cH\tensor\cF$ is the Hilbert space of an $N {=} 1$
superconformal field theory (sCFT) with $\hat c {=} 10$ or,
equivalently, $c {=} 15$.  This means that we can consider
$\cH\tensor\cF$ as a possible background for an $N{=}1$ string theory,
whose physical states will be given by the BRST cohomology
$H_{N{=}1}(\cH\tensor\cF)$.  In this language, the embedding of string
theories of \[BerVa] simply translates to
$$
H_{N{=}1}(\cH\tensor\cF) \isom H_{N{=}0}(\cH)~,\(ZeroInOne)
$$
where we mean an isomorphism as rings.  This point may need some
elaboration.  BRST cohomology has a natural ring structure: to every
BRST cocycle there corresponds a BRST invariant operator and the
OPE of such operators induce a multiplication on the physical states.
String amplitudes can in principle be computed from a knowledge of the
operator products of the physical operators, hence if two string
theories have isomorphic BRST cohomology rings, they are equivalent in
the sense that all amplitudes will coincide.  The equality between the
amplitudes was proven in \[BerVa] for those $\cH$ such that all the
cohomology is generated by ``matter'' excitations.  What we would like
to show in this note, is that \(ZeroInOne) is valid for all $\cH$.  We
will only partially succeed.  We will prove that \(ZeroInOne) is true
at the level of vector spaces.

The idea of the proof is very simple.  We approximate the computation
of $H_{N{=}1}(\cH\tensor\cF)$ by a spectral sequence whose second term
is precisely $H_{N{=}0}(\cH)$ and we show that this approximation is
actually exact.  The spectral sequence that will be useful in this
case arises out of a very obvious filtration of the complex computing
$H_{N{=}1}(\cH\tensor\cF)$.  This makes it possible to generalize this
result to other embeddings of string theories.  In an effort to keep
this note as short as possible, though, we will only discuss the
embedding \(ZeroInOne) and leave the more general results to a
lengthier and more detailed forthcoming publication.

The isomorphism \(ZeroInOne) is reminiscent of a well-known fact in
the theory of induced representations.  Suppose that $\gh \subset \gg$
are Lie algebras. Then there is a way to induce a representation of
$\gg$ from a representation of $\gh$, which goes by the name of the
induced module construction.  Indeed, if $V$ is any representation of
$\gh$ then it is a module over the universal enveloping algebra
$U(\gh)$.  We can then ``extend the scalars'' and define
$$
W \equiv U(\gg) \tensor_{U(\gh)} V ~.\(IndMod)
$$
$W$ is naturally a left $U(\gg)$-module and hence a representation of
$\gg$.  It is then a classic result (see, for example, \[Knapp]) that
in Lie algebra homology
$$
H_*(\gg ; W) \isom H_*(\gh ; V)~. \(HomIso)
$$
To determine whether \(ZeroInOne) is a semi-infinite instance of
\(HomIso) is beyond the scope of the present paper, but we will
explore this analogy further in the concluding section to see what it
can tell us about the search for the universal string theory.

\section{The Complexes}

Since we want to prove a result about the equality between two
cohomology spaces, we start by describing the complexes that compute
them.

As in the introduction, let $T_m(z)$ denote the energy momentum tensor
of a CFT with $c {=} 26$.  We will let $\cH$ denote the Hilbert space
of this CFT.  To compute $H_{N{=}0}(\cH)$ we introduce fermionic
ghosts $(b,c)$ of weights $(2,-1)$.  The BRST current is defined by
$$
J^{N{=}0}_{BRST} = T_m c + b c \d c \(jbrst)
$$
and its charge by
$$
Q_{N{=}0} = \contour{z}{0} J^{N{=}0}_{BRST}(z)~. \(qbrst)
$$
Because $c=26$ it follows that $Q_{N{=}0}^2=0$ whence its cohomology
$H_{N{=}0}(\cH)$ is well-defined.  It is known in the mathematical
literature as the semi-infinite cohomology of the Virasoro algebra
relative its center and with coefficients in the representation $\cH$.

We now introduce another fermionic BC system $(b_1,c_1)$ of weights
$(\fr3/2,-\half)$ and we call its Hilbert space $\cF$.  It carries a
representation of the Virasoro algebra with $c{=}-11$.  The tensor
product $\cH\tensor\cF$ carries therefore a representation of the
Virasoro algebra with $c{=}15$.  One of the remarkable results of
\[BerVa] is that it carries a representation of the $N{=}1$
(super)Virasoro algebra as well.  Indeed, $T$ and $G$ defined by
$$
\eqnalign{
T &\equiv T_m  - \fr3/2 b_1 \d c_1 - \half \d b_1 c_1 + \half \d^2(c_1\d
c_1)\(t)\cr
G &\equiv b_1 + c_1 T_m + c_1 \d c_1 b_1 + \fr5/2 \d^2 c_1 \(g)\cr}
$$
satisfy the OPEs defining the $N{=}1$ superconformal algebra with
$c=15$.  In other words, the tensor product $\cH\tensor\cF$ can be
understood as the Hilbert space of an $N{=}1$ sCFT with $c{=}15$.

Now, given any sCFT with $c{=}15$ and Hilbert space $\cS$, we can
define its BRST cohomology $H_{N{=}1}(\cS)$ as follows.  We introduce
in addition to the fermionic ghosts $(b,c)$ discussed above, a
bosonic BC system $(\beta,\gamma)$ of weights $(\fr3/2,-\fr1/2)$ and
we define the BRST current by
$$
J^{N{=}1}_{BRST} = T c + G \gamma + b c \d c - b \gamma^2 -
c\beta\d\gamma + \half \d c\beta\gamma~.\(jbrstI)
$$
Again because of $c{=}15$, its charge
$$
Q_{N{=}1} = \contour{z}{0} J^{N{=}1}_{BRST}(z) \(qbrstI)
$$
squares to zero and its cohomology---denoted by $H_{N{=}1}(\cS)$--can
be defined.

In particular when $\cS = \cH \tensor \cF$, the BRST current is given
by \(jbrstI) where $T$ and $G$ are given by \(t) and \(g)
respectively.  Explicitly, we find
$$
\eqnalign{
J^{N{=}1}_{BRST} ={}&
T_m c + b c \d c - b\gamma^2 + b_1\gamma - \fr3/2 b_1 \d c_1 c - b_1 \d
c_1 c_1 \gamma - \fr3/2 c\beta\d\gamma + T_m c_1 \gamma \cr
&-\fr1/2 \d b_1 c_1 c - \half c \d\beta\gamma + \fr5/2 \d^2c_1\gamma -
\half \d(\d^2 c_1 c_1) c~.\(expljbrst)\cr}
$$
It is this explicit form of $J^{N{=}1}_{BRST}$ that we shall exploit
to prove the isomorphism \(ZeroInOne).  We will see, however, that it
is rather the overall structure of $J^{N{=}1}_{BRST}$ that plays a
role and this makes possible the generalizations alluded to in the
introduction.

\section{The Spectral Sequence and the Calculation}

The isomorphism we are after essentially boils down to a cancellation
of the degrees of freedom of the fermionic BC system $(b_1,c_1)$ and
of the bosonic ghosts $(\beta,\gamma)$, which is reminiscent of the
quartet mechanism of Kugo and Ojima.  In fact, the original quartet
mechanism applied to abelian constraints.  In other words, it would
correspond to the BRST differential associated to the
current\fnote{The reason for the superscript will become obvious
shortly.} $J^{(0)} = b_1 \gamma$.  Interestingly enough, a glance at
the explicit expression \(expljbrst) for the $N{=}1$ BRST current
shows that such a term is indeed present.  The idea behind the proof
of \(ZeroInOne) is then to isolate this term by breaking up
$J^{N{=}1}_{BRST}$ into terms of different degrees:
$$
J^{N{=}1}_{BRST} = J^{(0)} + J^{(1)} +  \cdots \()
$$
in such a way that we can approximate the BRST cohomology of
$J^{N{=}1}_{BRST}$ by computing the cohomology induced by the
$J^{(i)}$.   The gadget that takes care of organizing this data is
called a spectral sequence and in the particular case that we are
considering, it will be the spectral sequence associated to a filtered
complex.

We choose to define the following degrees for our operators:
$$
\deg c = -\deg b = 1~, \quad \deg c_1 = \deg \gamma = - \deg b_1 = -
\deg\beta = 2~,\(degrees)
$$
and $\deg T_m = 0$.  Notice that this is compatible with the operator
product algebra.  Decomposing $J^{N{=}1}_{BRST}$ under this grading we
find that $J^{N{=}1}_{BRST} = \sum_{i=0}^5 J^{(i)}$ where $J^{(i)}$
has degree $i$ and they are given by
$$
\eqnalign{
J^{(0)} &= b_1 \gamma\cr
J^{(1)} &= T_m c + b c \d c - \fr3/2 b_1 \d c_1 c - \fr3/2 \beta\d\gamma c -
\fr1/2 \d b_1 c_1 c - \fr1/2 \d\beta\gamma c\cr
J^{(2)} &= 0\cr
J^{(3)} &= -b\gamma^2 - b_1 \d c_1 c_1 \gamma + T_m c_1 \gamma\cr
J^{(4)} &= \fr5/2 \d^2 c_1 \gamma\cr
\noalign{\hbox{and}}
J^{(5)} &= -\half \d (\d^2 c_1 c_1) c~.\cr}
$$
Let $Q_{N{=}1} = \sum_{i=0}^5 Q_{(i)}$ denote the decomposition of
the charge.  We can also break up the equation $Q_{N{=}1}^2=0$ into
components of different degrees.  Because $\deg$ is compatible with
the operator product algebra, each graded component has to be zero
separately and this means, in particular, that $Q_{(0)}^2 = 0$,
$[Q_{(0)},Q_{(1)}]=0$, and $Q_{(1)}^2 = 0$.  Among other things it
makes sense to talk about the cohomology of $Q_{(0)}$.

The above decomposition of $Q_{N{=}1}$ makes the BRST complex for the
$N{=}1$ string with that particular background into a filtered
complex.  Standard techinques in homological algebra now apply \[ss].
In particular, there exists a spectral sequence converging to the
cohomology of $Q_{N{=}1}$ and whose first term is the cohomology of
$Q_{(0)}$.  The space that $Q_{(0)}$ acts on is the tensor product of
the Hilbert spaces of the ``matter'' sCFT $\cH\tensor\cF$ and the
ghost sCFT $\cA_{b,c}\tensor\cS_{\beta,\gamma}$.  But since $Q_{(0)}$
does not depend on $(b,c)$ nor on $T_m$ it effectively acts only on
$\cF\tensor\cS_{\beta,\gamma}$.  Fortunately the cohomology of
$Q_{(0)}$ in this space has already been computed in \[homol].  In
fact, the Lemma in section 3 of \[homol] says that there is only one
nontrivial cocycle and this is the projective invariant vacuum.
Therefore we find that the first term in the spectral sequence---the
cohomology of $Q_{(0)}$---is isomorphic to $\cA_{b,c}\tensor\cH$.

The next term in the spectral sequence is the cohomology of $Q_{(1)}$
computed in the cohomology space of $Q_{(0)}$, that is, on
$\cA_{b,c}\tensor\cH$.   But on this space, $Q_{(1)}$ reduces to
$Q_{N{=}0}$ and hence the second term in the spectral sequence is
$H_{N{=}0}(\cH)$.

All the other $Q_{(i>1)}$ are automatically zero since they involve
the $(\beta,\gamma)$ and $(b_1,c_1)$ fields in one way or another.
Therefore we conclude that the spectral sequence degenerates at the
second term yielding the isomorphism \(ZeroInOne).  Notice that the
proof does not rely on the particular form of $T_m$ hence it is true
for arbitrary $N{=}0$ string backgrounds.  The proof has one
shortcoming, though.  It does not respect the ring structure.  And
although it seems rather plausible that \(ZeroInOne) does indeed hold
as rings, we cannot unfortunately conclude this from our proof.

\section{On the Concept of a Universal String Theory}

As mentioned in the introduction, the isomorphism \(ZeroInOne) is
reminiscent of the analogous result in the theory of Lie algebra
homology by which to compute the homology of a Lie algebra $\gh$
with coefficients in a representation $V$, one can proceed directly
or, alternatively, one can first embed $\gh$ in $\gg$ and then compute
the homology of $\gg$ with coefficients in the representation induced
by $V$---the point being that all the homological information of the
induced representation is contained already in $V$.

In our case, the Virasoro algebra plays the role of $\gh$, the $N{=}1$
supervirasoro algebra plays the role of $\gg$ and tensoring with the
CFT of the $(b_1,c_1)$ system seems to play the role of the induced
module construction \(IndMod).  It is not inconceivable that this
analogy contains some element of truth and that this construction is
in fact a semi-infinite version of the induced module construction,
but we will not attempt to elucidate this point here and
now.\fnote{Although see \[Sasha] for a semi-infinite version of
\(HomIso).  It would be interesting to see if both constructions
coincide for this case.} Nevertheless, let us pursue the analogy a
little bit farther to see what it implies on the existence of the
universal string theory.

What seems to emerge is that there is not a unique universal string
theory, but rather that there is probably a universal string theory in
each hierarchy of embeddings we can consider.  For example, in the Lie
algebra case, there are many chains of embeddings into which, say, $sl_2$
fits.  For example we can take the following
$$
sl_2 \subset sl_3 \subset sl_4 \subset \cdots \(chainI)~.
$$
The functorial nature of the induced module construction guarantees
that the two ways we can induce a representation in $sl_4$, say,
starting from a representation of $sl_2$---that is, either by
considering the embedding $sl_2\subset sl_4$ directly, or else by
going through $sl_3$---are actually equivalent.  Therefore we can take
the inductive limit of \(chainI)---call it $sl_\infty$---and induce a
module from a module in $sl_2$.  However notice that depending on how
we actually embed $sl_n$ in $sl_{n+1}$ we will get a different
$sl_\infty$ and thus a different ``universal'' algebra.  Worse yet, we
could make use of the accidental isomorphism of complex Lie algebras
$sl_2 \isom so_3$ to consider other chains of embeddings
like
$$
sl_2 \isom so_3 \subset so_5 \subset so_7 \subset \cdots~, \(chainII)
$$
which would again lead to a different universal object.

Similarly in string theory, we could choose as in \[BerVa] to embed
the $N{=}0$ Virasoro algebra in the $N{=}1$ superVirasoro algebra and
this in turn inside the $N{=}2$ superVirasoro algebra, but one could
equally well have started by embedding it in, say, $\W_3$ and pursue
another chain of embeddings.\fnote{Note, however, that there is no
chain of embeddings associated to the $\W_n$ algebras, but rather (at
least in the classical case) an inverse system of reductions: $\W_n
\leftarrow \W_{n+1}$. (This was exploited in \[uniw] to define the
notion of a universal $\W$-algebra.)  But we mean any chain of
embeddings of increasingly more symmetrical CFTs, the first two of
which have as chiral algebras the Virasoro algebra and $\W_3$
respectively.}  Presumably both directions would yield different
universal string theories.  Why one would choose one direction in
which to embed a particular string theory over any other seems to be
purely a phenomenological matter; but then again, the choice of vacuum
seems to have the final word on the phenomenology.  It would seem
therefore that we need some more clues to narrow the search for the
universal string theory.

\ack

It's a pleasure to thank Chris Hull for discussions on universal
string theory; Nathan Berkovits and Cumrun Vafa for valuable
comments on a previous version of this paper, and Alexander Voronov
for pointing out his semi-infinite version of \(HomIso).
\refsout
\bye